# Unexpected Delta-Function Term in the Radial Schrodinger Equation

Anzor A.Khelashvili and Teimuraz P. Nadareishvili

*Inst. of High Energy Physics, Iv. Javakhishvili Tbilisi State University, University Str. 9, 0109, Tbilisi, Georgia*
*and St.Andrea the First-called Georgian University of Patriarchy of Georgia, Chavchavadze Ave.53a, 0162, Tbilisi, Georgia.*

**E-mail:** teimuraz.nadareishvili@tsu.ge and anzor.khelashvili@tsu.ge

**Abstract:** Careful exploration of the idea that equation for radial wave function must be compatible with the full Schrodinger equation shows appearance of the delta-function while reduction of full Schrodinger equation in spherical coordinates. Elimination of this extra term produces a boundary condition for the radial wave function, which is the same both for regular and singular potentials.

1. *Introduction.* - It is well known that the radial Schrodinger equation

$$-\frac{d^2 u(r)}{dr^2} + \frac{l(l+1)}{r^2} u(r) - 2m[E - V(r)] u(r) = 0 \qquad (1)$$

plays a central role in quantum mechanics due to frequent encounter with the spherically symmetric potentials. In turn, this equation is obtained from the full 3-dimensional Schrodinger equation

$$-\Delta \psi(\vec{r}) - 2m[E - V(r)]\psi(\vec{r}) = 0 \qquad (2)$$

after the separation of variables in spherical coordinates [1, 2].

Recently considerable attention has been devoted to the problems of self-adjoint extension (SAE) for the inverse squared $r^{-2}$ behaved potentials in the radial Schrodinger equation [3]. These problems are interesting not only from academic standpoint, but also due to large number of physically significant quantum-mechanical problems that manifest in such a behavior.

Hamiltonians with inverse squared like potentials appear in many systems and they have sufficiently rich physical and mathematical structures. Starting from 60-ies of the previous century, singular potentials were the subject of intensive studies in connection with non-renormalizable field theoretic models. Exhaustive reviews dedicated to singular potentials for that era can be found in [4-6].

It turned out that there are no rigorous ways of deriving some boundary condition for the radial wave function $u(r)$ from the radial equation itself at the origin $r = 0$ in case of singular potentials.

Therefore, many authors content themselves by consideration only a square integrability of radial wave function and do not pay attention to its behavior at the origin. Of course this is permissible mathematically and the strong theory of linear differential operators allows for such approach [7-8]. There appears so-called SAE physics [3], in the framework of which among physically reasonable solutions one encounters also many curious results, such as bound states in case of repulsive potential [9] and so on. We think that these highly unphysical results are caused by the fact that without suitable boundary condition at the origin a functional domain for radial Schrodinger Hamiltonian is not restricted correctly [10].

The aim of this letter is to establish the boundary condition for the radial wave function $u(r)$. Below we show that, owing to the singular character of transformation leading to Eq. (1) from Eq. (2), there appears extra delta function term, which plays a role of point-like source, interacting with the wave function. Surprisingly enough, this term has not been noted earlier. From the requirement of its absence definite boundary condition follows on the radial wave function at the origin. This fact can have a great influence on the further considerations of the radial equation.

2. *Rigorous derivation of boundary condition.* -Let us mention, that the transition from Cartesian to spherical coordinates is not unambiguous, because the Jacobian of this transformation $J = r^2 \sin\theta$ is singular at $r = 0$ and $\theta = n\pi (n = 0,1,2,...)$. Angular part is fixed by the requirement of continuity and uniqueness. This gives the unique spherical harmonics $Y_l^m(\theta,\varphi)$.

We also note in regards to radial variable that, although $\vec{r} = 0$ is an ordinary point in full Schrodinger equation, it is a point of singularity in the radial equation and thus, knowledge of specific boundary behavior is necessary.

We have to bear in mind that the radial Eq.(1) is not independent equation but is derived from full 3-dimensional Schrodinger equation (2) and as it is underlined in many classical books on quantum mechanics, the final radial equation must be compatible with the primary full Schrodinger equation. Unfortunately, in our opinion, this consideration has not been extended to any concrete results [2, 11]. Though several discussions of mostly "beat around the bush" nature exist in the literature (see, e.g. book of R. Newton [12]), the conclusions from these studies are largely conservative and cautious. It seems that without deeper exploration of the idea of compatibility, some significant point will be missing.

Armed with this idea let us now look at derivation of the radial wave equation in more details. Remembering that, after substitution

$$\psi(\vec{r}) = R(r) Y_l^m(\theta,\varphi) \qquad (3)$$

into the 3-dimensional Equation (2), it follows the usual form of equation for full radial function $R(r)$:

$$\frac{d^2 R}{dr^2} + \frac{2}{r}\frac{dR}{dr} + 2m[E - V(r)]R - \frac{l(l+1)}{r^2}R = 0 \qquad (4)$$

It is a common practice to avoid the first derivative term from this equation by substitution

$$R(r) = \frac{u(r)}{r} \qquad (5)$$

Because this substitution enhances singularity at $r = 0$, care must be taken to perform it. Let us rewrite equation (4) after using Eq. (5)

$$\frac{1}{r}\left(\frac{d^2}{dr^2} + \frac{2}{r}\frac{d}{dr}\right)u(r) + u(r)\left(\frac{d^2}{dr^2} + \frac{2}{r}\frac{d}{dr}\right)\left(\frac{1}{r}\right) + 2\frac{du}{dr}\frac{d}{dr}\left(\frac{1}{r}\right) + \left[-\frac{l(l+1)}{r^2} + 2m(E - V(r))\right]\frac{u}{r} = 0 \qquad (6)$$

We write equation in this form deliberately, to indicate action of radial part of Laplacian on relevant factors explicitly. It seems that the first derivatives of $u(r)$ cancelled out and yielding following equation

$$\frac{1}{r}\left(\frac{d^2 u}{dr^2}\right) + u\left(\frac{d^2}{dr^2} + \frac{2}{r}\frac{d}{dr}\right)\left(\frac{1}{r}\right) - \frac{l(l+1)}{r^2}\frac{u}{r} + 2m(E - V(r))\frac{u}{r} = 0 \qquad (7)$$

If we now differentiate the second term "naively", we'll derive zero. But such treatment would be valid only in the case, when $r \neq 0$. However, in general this term is proportional to the 3-dimensional delta function. Indeed noticing that

$$\frac{d^2}{dr^2} + \frac{2}{r}\frac{d}{dr} = \frac{1}{r^2}\frac{d}{dr}\left(r^2 \frac{d}{dr}\right) \equiv \Delta_r$$

is the radial part of the Laplace operator and following [13]

$$\Delta_r\left(\frac{1}{r}\right) = \Delta\left(\frac{1}{r}\right) = -4\pi\delta^{(3)}(\vec{r}) \tag{8}$$

we obtain the equation for $u(r)$

$$\frac{1}{r}\left[-\frac{d^2 u(r)}{dr^2} + \frac{l(l+1)}{r^2}u(r)\right] + 4\pi\delta^{(3)}(\vec{r})u(r) - 2m[E - V(r)]\frac{u(r)}{r} = 0 \tag{9}$$

Thus, there appears the extra delta-function term, which must be eliminated. Note that when $r \neq 0$, this extra term vanishes owing to the property of the delta function and, in this case, if we multiply equation (9) on $r$, we obtain the ordinary radial equation (1).

However if $r = 0$, multiplication on r is not permissible and this the extra delta-function term remains in Eq.(9) Therefore one has to investigate this term separately and find another way to eliminate it.

The term with 3-dimensional delta-function must be comprehended as being integrated over $d^3r = r^2 dr \sin\theta d\theta d\varphi$. On the other hand [13],

$$\delta^{(3)}(\vec{r}) = \frac{1}{|J|}\delta(r)\delta(\theta)\delta(\varphi) \tag{10}$$

Taking into account all the above mentioned relations, it is clear that the extra term still survives, but now in the one-dimensional form

$$u(r)\delta^{(3)}(\vec{r}) \to u(r)\delta(r) \tag{11}$$

Its appearance as a point-like source breaks many fundamental principles of physics, which is not desirable. The only reasonable way to remove this term without modifying Laplace operator or including compensating delta function term in the potential $V(r)$, is the requirement

$$u(0) = 0 \tag{12}$$

(note, that multiplication of Eq. (9) on $r$ and then elimination this extra term owing the property $r\delta(r) = 0$ is not legitimate procedure, because effectively it is equivalent to multiplication on zero).

Therefore we conclude that the radial equation (1) for $u(r)$ is compatible with the full Schrodinger equation (2) if and only if the boundary condition $u(0) = 0$ is fulfilled. *The radial equation (1) supplemented by the boundary condition (12) is equivalent to the full Schrodinger equation (2).* It is in accordance with the Dirac requirement, that the solutions of the radial equation must be compatible with the full Schrodinger equation [2].

3. *Comments, some applications and conclusions* - Some comments are in order here:
The main conclusion of our work is that traditional radial equation must be necessarily supplemented by boundary condition $u(0) = 0$, otherwise this equation does not take place. Therefore obtained result is not only mathematical novelty, but it also ensures the usual status of the radial equation, which has been the main quantum-mechanical equation during last 90 years. Owing to this fact there are at least two important new conclusions - one positive and one negative.

Positive news is that, papers in which this boundary condition has been used are correct. But papers without this boundary condition are wrong (there have been more than hundred such papers published, including publications in PRL. Part of them is cited in [10]).

Moreover in all significant books on quantum mechanics, beginning from the classical ones and including modern textbooks, there are errors in formulation of the radial equation..

It is also remarkable to note that the boundary condition (12) is valid whether potential is regular or singular. It is only consequence of the particular transformation of the Laplacian. Different potentials can only determine the specific way of $u(r)$ tending to zero at the origin and the delta function arises in the reduction of the Laplace operator every time. All of these statements can easily be verified also by explicit integration of Eq. (9) over a small sphere with radius $a$ tending it to zero at the end of calculations.

It seems very curious that this fact has been unnoticed up till now in spite of numerous discussions [2,5,6,11,12]. Now, when this boundary condition has been established, many problems can be solved by taking it into account. Remarkably, all the results obtained earlier for regular potentials with the boundary condition (12) remain almost unchanged. In majority of textbooks on quantum mechanics, $r \to 0$ behavior is obtained from Eq. (1) in case of regular potentials. But we have shown that this equation takes place only together with boundary condition (12). On the other hand, for *singular potentials* this condition will have far-reaching implications. Many authors neglected boundary condition entirely and were satisfied only by square integrability. But this treatment, after leakage into the forbidden regions and through a self-adjoint extension procedure, sometimes yields curious unphysical results. Below we consider some simple examples, showing the differences, which arise with and without above mentioned boundary condition:

(i) regular potentials:
$$\lim_{r \to 0} r^2 V(r) = 0$$

In this case, after substitution at the origin $u \sim r^a$, indicial equation follows $a(a-1) = l(l+1)$, giving two solutions $u \underset{r \to 0}{\sim} c_1 r^{l+1} + c_2 r^{-l}$ (see, any textbooks on quantum mechanics). For non-zero $l$-s the second solution is not square integrable and is ignored usually. But for $l=0$, many authors [14-15] discuss how to deal with this solution, which is also square integrable at the origin. According to our result, this solution must be ignored.

(ii) transitive singular potentials:
$$\lim_{r \to 0} r^2 V(r) = -V_0 = const$$

In this case the indicial equation takes form $a(a-1) = l(l+1) - 2mV_0$, which has two solutions: $a = \frac{1}{2} \pm \sqrt{\left(l+\frac{1}{2}\right)^2 - 2mV_0}$. Therefore

$$u \underset{r \to 0}{\sim} c_1 r^{\frac{1}{2}+P} + c_2 r^{\frac{1}{2}-P} \; ; \quad P = \sqrt{\left(l+\frac{1}{2}\right)^2 - 2mV_0} \tag{13}$$

It seems, that both solutions are square integrable at origin as long as $0 \leq P < 1$. Precisely this range is studied in most papers (See relevant citations in [10]), whereas according to our boundary condition we have $0 \leq P < \frac{1}{2}$. The difference is essential. Indeed, the radial equation the has form

$$u'' - \frac{P^2 - 1/4}{r^2} u = 2mE$$

Depending on whether $P$ exceeds 1/2 or not, the sign in front of the fraction changes and one can derive attraction in case of repulsive potential and vice versa. Boundary condition avoids these unphysical solutions.

Lastly, we note that the same holds for radial reduction of the Klein-Gordon equation, because in three dimensions it has the following form

$$(-\Delta + m^2)\psi(\vec{r}) = [E - V(r)]^2 \psi(\vec{r}) \tag{14}$$

and the reduction of variables in spherical coordinates will proceed to absolutely same direction as in Schrodinger equation.

We want to thank Profs. Sasha Kvinikhidze and Parmen Margvelashvili for valuable discussions. A.Kh. is indebted to thank Prof. Boris Arbuzov for reading the manuscript.